\documentclass[12pt]{article} 
\usepackage{psfig} 
 
  \advance\oddsidemargin by 15mm 
  \advance\evensidemargin by 15mm 
  \advance\topmargin by 10mm 
 
\begin{document} 
 
\noindent 
\begin{minipage}[t][6cm][t]{18cm} 
 
{\large\bf Temperature evolution of the quantum Hall effect in 
the FISDW state: \\ 
Theory vs Experiment}\\ 
 
\bigskip 
 
Victor M. Yakovenko and Hsi-Sheng Goan \\ 
{\it Department of Physics, University of Maryland, College Park, MD 
20742-4111, USA}\\ 
 
\bigskip 
 
Jonghwa Eom and Woowon Kang \\ 
{\it James Franck Institute and Department of Physics, 
University of Chicago, Chicago, IL 60637, USA} 
 
\end{minipage} 
 
\noindent 
\hspace*{15mm}\parbox[t]{16.5cm} {\small 
 
  {\bf Abstract.} We discuss the temperature dependence of the Hall 
  conductivity $\sigma_{xy}$ in the magnetic-field-induced 
  spin-density-wave (FISDW) state of the quasi-one-dimensional 
  Bechgaard salts (TMTSF)$_2$X.  Electronic thermal excitations across 
  the FISDW energy gap progressively destroy the quantum Hall effect, 
  so $\sigma_{xy}(T)$ interpolates between the quantized value at zero 
  temperature and zero value at the transition temperature $T_c$, 
  where FISDW disappears.  This temperature dependence is similar to 
  that of the superfluid density in the BCS theory of 
  superconductivity.  More precisely, it is the same as the 
  temperature dependence of the Fr\"ohlich condensate density of a 
  regular CDW/SDW.  This suggests a two-fluid picture of the quantum 
  Hall effect, where the Hall conductivity of the condensate is 
  quantized, but the condensate fraction of the total electron density 
  decreases with increasing temperature.  The theory appears to agree 
  with the experimental results obtained by measuring all three 
  components of the resistivity tensor simultaneously on a 
  (TMTSF)$_2$PF$_6$ sample and then reconstructing the conductivity 
  tensor.} 
 
\vspace{2cm} 
 
In a magnetic field $H$, quasi-one-dimensional organic conductors of 
the (TMTSF)$_2$X family experience a cascade of magnetic-field-induced 
spin-density-wave (FISDW) phase transitions (see review 
\cite{Chaikin96}).  At zero temperature, each FISDW phase exhibits an 
integer quantum Hall effect (see review \cite{Yakovenko96c}).  As 
temperature $T$ increases, the Hall effect decreases and virtually 
vanishes at $T_c$, the transition temperature of FISDW.  (In the 
normal state without FISDW, the Hall effect is very small and can be 
approximated as zero compared to the Hall effect in the FISDW state.) 
It was shown in Refs.\ \cite{Yakovenko96c,Yakovenko97b,Yakovenko98c} 
that the temperature evolution of the Hall conductivity (per one 
layer) is given by the following formula: 
\begin{equation} 
\sigma_{xy}(T)=f(T)\,\frac{2Ne^2}{h}, 
\label{sigma_xy} 
\end{equation} 
where $e$ is the electron charge, $h$ is the Planck constant, $N$ is 
an integer number that characterizes the FISDW, and $f(T)$ is the 
dimensionless condensate density of FISDW.  The condensate density 
$f(T)$ interpolates between 1 at zero temperature (all electrons are 
in the condensate) and 0 at $T_c$ (the condensate density vanishes 
when FISDW disappears).  An explicit expression for $f(T)$ depends on 
the order in which the limits of zero frequency $\omega\to0$ and zero 
momentum $q\to0$ are taken \cite{Lee79,Maki90}.  In the dynamic limit, 
where the $q\to0$ is taken first and then $\omega\to0$ is taken, the 
condensate density is 
\begin{equation} 
f_d(T)=1-\int_{-\infty}^\infty \frac{dp_x}{v_{\rm F}} 
\left(\frac{\partial E}{\partial p_x}\right)^{\!2} 
\left[-\frac{\partial n_{\rm F}(E)}{\partial E}\right], 
\label{fd} 
\end{equation} 
where $E=\sqrt{(v_{\rm F}p_x)^2+\Delta^2}$ is the electron energy 
dispersion law in the FISDW state ($v_F$ is the Fermi velocity, $p_x$ 
is the electron momentum along the chains, and $\Delta$ is the energy 
gap), and $n_{\rm F}(\epsilon)=(e^{\epsilon/k_{\rm B}T}+1)^{-1}$ is 
the Fermi distribution function ($k_{\rm B}$ is the Boltzmann 
constant).  In the static limit ($\omega\to0$ first, then $q\to0$), 
the condensate density is 
\begin{equation} 
f_s(T)=1-v_{\rm F}\int_{-\infty}^\infty dp_x 
\left[-\frac{\partial n_{\rm F}(E)}{\partial E}\right]. 
\label{fs} 
\end{equation} 
The static limit (\ref{fs}) is appropriate for calculation of the 
magnetic field penetration depth in superconductors, which comes from 
the truly static Meissner effect in thermodynamic equilibrium.  On the 
other hand, the Hall effect is kinetic and not thermodynamic, so we 
believe that the dynamic limit (\ref{fd}) is appropriate for Eq.\ 
(\ref{sigma_xy}). 
 
\begin{figure} 
\centerline{\psfig{file=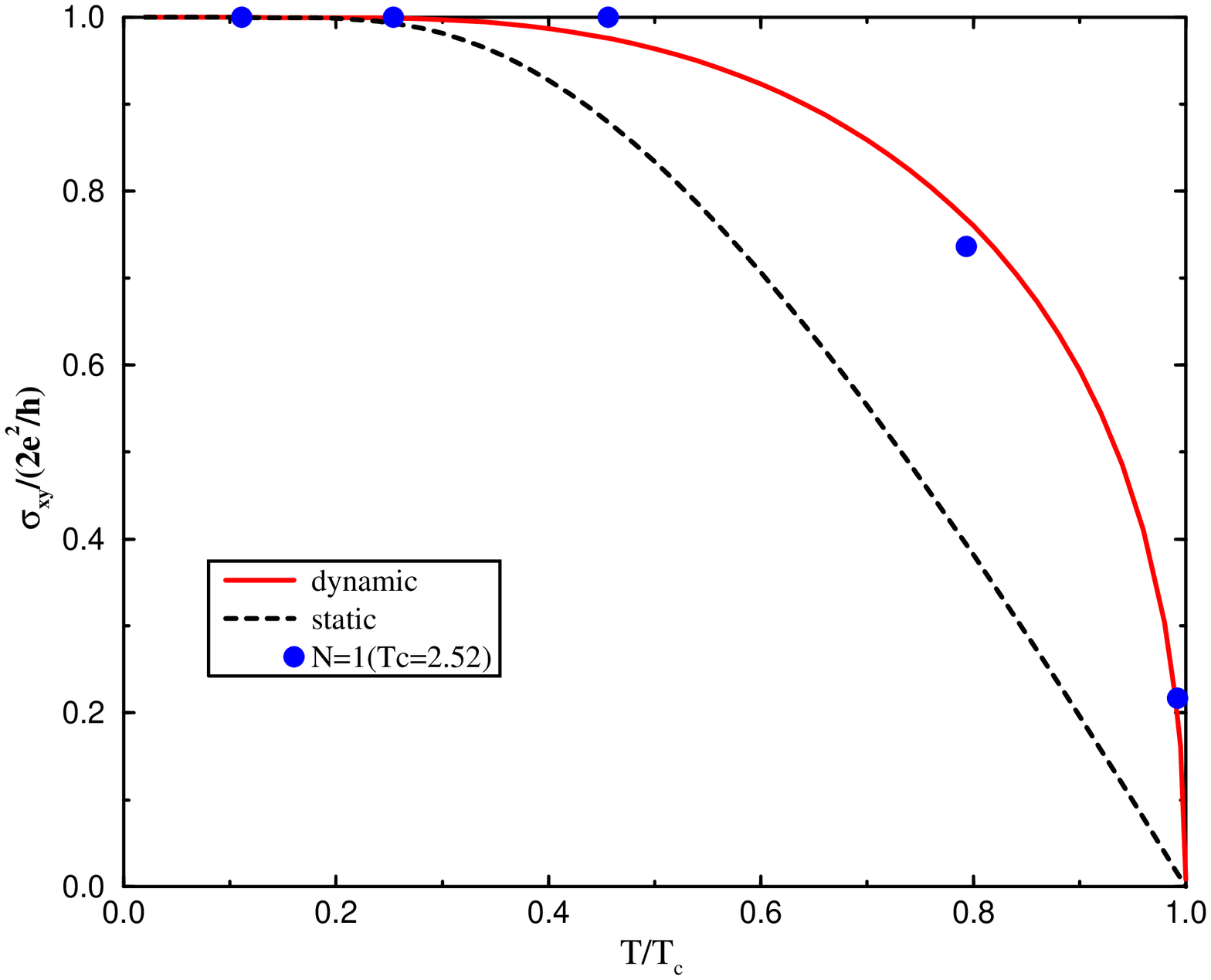,width=0.49\textwidth,angle=-90} 
\hfill 
\psfig{file=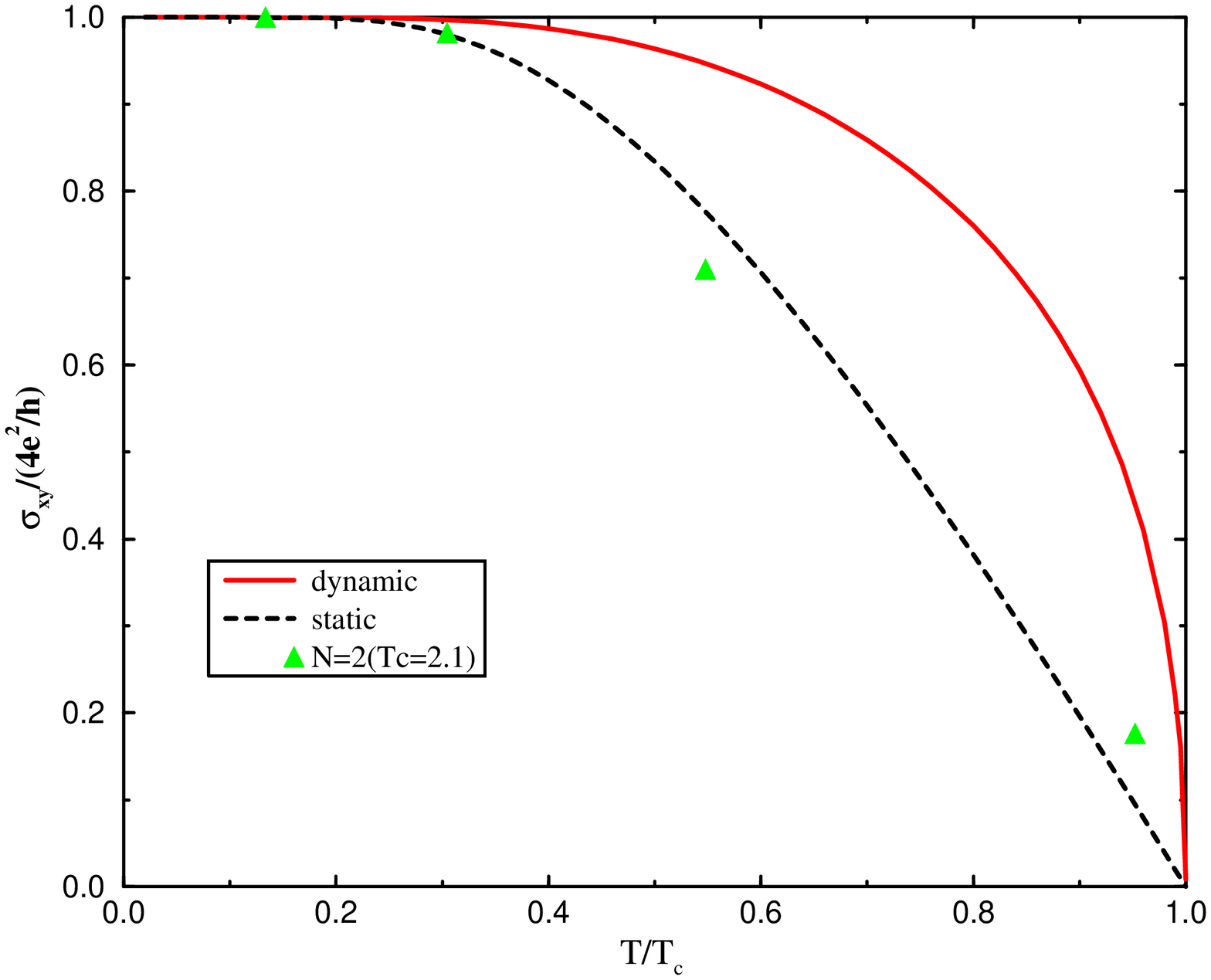,width=0.49\textwidth,angle=-90}} 
\caption{\small 
  Hall conductivity $\sigma_{xy}$ normalized to its zero-temperature 
  value as a function of temperature normalized to the transition 
  temperature of FISDW.  The solid and dashed lines represent the 
  temperature dependences of the condensate density in the dynamic and 
  static limits, respectively, calculated from Eqs.\ (\ref{fd}) and 
  (\ref{fs}).  The left panel shows experimental points for 
  (TMTSF)$_2$PF$_6$ under pressure 8.5 kbar for the $N=1$ Hall plateau 
  at $H=15.5$ T assuming $T_c=2.52$ K, and the right panel for $N=2$ at 
  $H=13.25$ T assuming $T_c=2.1$ K.} 
\label{figure} 
\end{figure} 
 
In Fig.\ \ref{figure}, we compare theory and experiment.  Hall 
conductivity $\sigma_{xy}$ normalized to its zero-temperature value is 
shown as a function of temperature normalized to the transition 
temperature of FISDW.  The solid and dashed lines represent the 
temperature dependences of the condensate density in the dynamic and 
static limits, respectively, calculated from Eqs.\ (\ref{fd}) and 
(\ref{fs}) presuming that the temperature dependence of the energy gap 
$\Delta$ is the same as in the BCS theory of superconductivity 
\cite{Montambaux86}.  The points are obtained experimentally by 
measuring the three components of the resistivity tensor, $\rho_{xx}$, 
$\rho_{yy}$, and $\rho_{xy}$, simultaneously on a sample of 
(TMTSF)$_2$PF$_6$ under pressure 8.5 kbar and then calculating the 
conductivity tensor.  The left panel shows the experimental points for 
the $N=1$ Hall plateau at $H=15.5$ T assuming $T_c=2.52$ K, and the 
right panel for $N=2$ at $H=13.25$ T assuming $T_c=2.1$ K.  The left 
panel demonstrates a good agreement with the dynamic limit.  The right 
panel seems to agree with the static limit.  However, the right panel 
has only four experimental points, and the discrimination between the 
static and dynamic fits is effectively controlled by only one point at 
about $T/T_c=0.6$.  So, we believe that there is not enough data to 
make a conclusion for the $N=2$ plateau, but for the $N=1$ plateau 
the agreement with the dynamic limit appears to be convincing. 
Nevertheless, more data is necessary to make a firmer 
conclusion.  We also 
wish to point out that Eq.\ (\ref{sigma_xy}) provides a unique 
opportunity to determine the condensate density of FISDW by measuring 
the Hall conductivity.  This is a linear-response measurement, 
uncomplicated by depinning, phase slips, and other nonlinear effects, 
which characterize sliding of a regular charge- or spin-density wave. 
 
The last terms in Eqs.\ (\ref{fd}) and (\ref{fs}) reflects the fact 
that normal quasiparticles, thermally excited above the energy gap, 
reduce the quantum Hall effect.  We can write the condensate density 
in the form 
\begin{equation} 
f(T)=1-\rho_n/\rho, 
\label{f(T)} 
\end{equation} 
where the second term is the density of the normal component, 
$\rho_n$, normalized to the total electron density $\rho$.  Thus, 
$f(T)$ is analogous to the superfluid density in superconductors. 
Given that normal electrons exhibit virtually zero Hall effect, Eq.\ 
(\ref{sigma_xy}) can be interpreted in a two-fluid manner: The 
condensate carries the dissipationless quantum Hall current, and 
normal electrons carry the longitudinal, dissipative electric current. 
The Hall response of the condensate remains quantized even at a finite 
temperature, but the Hall conductivity decreases because the 
condensate density decreases with increasing temperature.  In this 
picture, the quantum Hall current in the FISDW state is analogous to 
the supercurrent in a superconductor, which is also carried only by a 
condensate, whose density decreases with increasing temperature.  This 
analogy continues an amazing sets of parallels between the quantum 
Hall effect in the FISDW state and superconductivity.  The 
thermodynamics of both states is described by the BCS theory of a 
mean-field energy gap $\Delta$ opening at the Fermi surface 
\cite{Montambaux86}.  Both states are characterized by dissipationless 
currents: the quantum Hall current and the supercurrent.  At a finite 
temperature, both systems have two-fluid electrodynamics. 
 
The temperature dependence discussed in this paper was obtained under 
assumption that the quantum Hall effect originates from the bulk of 
the crystal.  The results have implications for the edge states 
picture of the quantum Hall effect.  Theory says that in the FISDW 
state there must be $N$ gapless chiral electron states at the sample 
edge \cite{Yakovenko96c}.  At zero temperature, the quantized Hall 
conductivity can be equivalently obtained in terms of either bulk or 
edge states \cite{Yakovenko96c}.  However, it does not seem possible 
to describe the Hall conductivity at a finite temperature in terms of 
only the edge states.  Indeed, the edge states are gapless, so they 
cannot produce Eqs.\ (\ref{fd}) and (\ref{fs}), which involve the gap 
$\Delta$.  Moreover, because of thermal excitations across the energy 
gap in the bulk, it does not seem possible to ignore the bulk 
contribution compared to the edge contribution at a finite 
temperature.  By any means, at $T\to T_c$ distinction between the bulk 
and edge states must go away.  So, the bulk theory of the quantum Hall 
effect appears to be more general and robust than the edge one.  It is 
worth mentioning that the question where the quantum Hall current 
actually flows, in the bulk or in the edges, was never studied 
experimentally in quasi-one-dimensional conductors and does not have a 
clear answer even in the case of the quantum Hall effect in 
semiconductors. 
 

\end{document}